\documentclass[twocolumn,showpacs,preprintnumbers,amsmath,amssymb]{revtex4}
\usepackage{graphicx}
\usepackage{dcolumn}
\usepackage{bm}
\newcommand{\vev}[1]{\langle {#1}\rangle}

\def\e6{E(6)}
\def\10{SO(10)}
\def\21{SU(2) $\otimes$ U(1) }

\def\3lrBL{$SU(3)_{c}\otimes SU(2)_L \otimes SU(2)_R \otimes U(1)_{B-L}$}
\def\422{SU(4) $\otimes$ SU(2) $\otimes$ SU(2)}
\newcommand{\AHEP}{%
  AHEP Group, Instituto de F\'{\i}sica Corpuscular --
  C.S.I.C./Universitat de Val{\`e}ncia \\
  Edificio Institutos de Paterna, Apt 22085, E--46071 Valencia, Spain}
\newcommand{\Lisb}{%
 Departamento de F\'\i sica and CFTP, Instituto Superior T\'ecnico\\
          Av. Rovisco Pais 1, $\:\:$ 1049-001 Lisboa, Portugal }
\newcommand{\SISSA}{%
Scuola  Superiore
di Studi Avanzati\\
Via Beirut 4, I-34014, 
and INFN, Sezione di Trieste, Italy}
\def\roughly#1{\mathrel{\raise.3ex\hbox{$#1$\kern-.75em
      \lower1ex\hbox{$\sim$}}}} 

\include{definitions}                                                                                
\begin{document}
\preprint{IFIC/05-28}
\title{Novel Supersymmetric SO(10) Seesaw Mechanism}
\date{\today}
\author{M. Malinsk\'y}\email{malinsky@sissa.it}
\affiliation{\SISSA}
\author{J.C.Rom\~ao}\email{jorge.romao@ist.utl.pt}
\affiliation{\Lisb}
\author{J.W.F.Valle}\email{valle@ific.uv.es}
\affiliation{\AHEP}
\begin{abstract}
  
  We propose a new seesaw mechanism for neutrino masses within a class
  of supersymmetric SO(10) models with broken D-parity. It is shown
  that in such scenarios the $B-L$ scale can be as low as TeV without
  generating inconsistencies with gauge coupling unification nor with
  the required magnitude of the light neutrino masses. This leads to a
  possibly light new neutral gauge boson as well as relatively light
  quasi-Dirac heavy leptons. These particles could be at the TeV scale
  and mediate lepton flavour and CP violating processes at appreciable
  levels.

 \end{abstract}
\maketitle


The origin of neutrino masses is the most well kept secret of modern
elementary particle physics. The basic dimension--five operator which
leads to neutrino masses~\cite{Weinberg:1980bf} can arise from physics
at vastly different scales. One popular alternative is the seesaw
mechanism, in which case the small neutrino masses are induced by the
exchange of superheavy neutral
fermions~\cite{Minkowski:1977sc,gell-mann:1980vs,yanagida:1979,Mohapatra:1980ia}
or superheavy scalars, or
both~\cite{schechter:1980gr,mohapatra:1981yp,Lazarides:1980nt}.  The
light neutrino masses are given as
\begin{equation}
 \label{eq:nss}
M_{\nu} \simeq - v^{2} Y {M}^{-1} Y^{T}
\end{equation}
An alternative inverse seesaw scheme has been
suggested~\cite{mohapatra:1986bd} for theories which lack the
representation required to implement the canonical seesaw, as happens
in a class of string inspired models. 


In addition to the normal neutrinos $\nu$ such inverse seesaw
mechanism employs two sequential SU(3) $\otimes$ SU(2) $\otimes$ U(1)
singlets $\nu^{c}$, $S$ (these are all left-handed
two-component spinors~\cite{schechter:1980gr}).  The effective
neutrino mass matrix has the following form:
\begin{equation}
\label{generalM} M_{\nu} =
\left(\begin{array}{ccc}
    0 & Y v& 0 \\
    Y^{T}v & 0 & M\\
    0 & M & \mu
\end{array}\right) 
\end{equation}
in the basis $\nu_{L}, \nu^{c}_{L}$, $S_{L}$. Here $Y$ is the Yukawa
matrix parametrizing the $Y^{ij}
{\nu_{L}^i}^{T}C^{-1}{\nu^j}^{c}_{L}+h.c.$ interactions, while $M$ and
$\mu$ are SU(3) $\otimes$ SU(2) $\otimes$ U(1) invariant mass entries.
When $\mu \to 0$ a global lepton--number symmetry is exactly conserved
and all three light neutrinos are strictly massless. Yet it has been
shown~\cite{bernabeu:1987gr,Ilakovac:1994kj} that lepton flavour and
CP can be violated at appreciable levels even in the absence of
supersymmetry, provided the scale $M$ is sufficietly low.
When $\mu \neq 0$ the mass matrix for the light eigenstates is given
by
\begin{equation}
 \label{eq:iss}
M_{\nu}\simeq - v^{2}(Y {M^{T}}^{-1}) \mu ({M}^{-1}Y^{T})
\end{equation}
One sees that neutrinos can be made very light, as required by
oscillation data~\cite{Maltoni:2004ei}, even if $M$ is very low, far
below the GUT scale $M_{G}$ ($M \ll M_{G}$), provided $\mu$ is very small,
$\mu \ll M$.  This scheme has a very rich and interesting
phenomenology, since no new scales need to be added to generate the
small neutrino masses, instead a small parameter $\mu$ is added.  Note
that in such SU(3) $\otimes$ SU(2) $\otimes$ U(1) inverse seesaw the
smallness of $\mu$ is natural, in t'Hooft's
sense~\cite{'tHooft:1979bh}, as the symmetry enhances when $\mu \to
0$. However, there is no dynamical understanding of this smallness.

Here we provide an alternative inverse seesaw realization consistent
with a realistic unified \10 model. Such embedding brings several
issues:
\begin{itemize}
\item{The $\nu^c S$ entry $M$ (generated by the VEV of a Higgs multiplet $\chi_{R}$ 
    with \3lrBL quantum numbers $(1,1,2,-1)$) breaks the $B-L$ symmetry, now gauged.
    The corresponding scale $\vev{\chi_{R}}$ must be compatible with gauge
    coupling unification.  Together with the requirement of 
    low-energy supersymmetry to stabilize the hierarchies, this places
    rather strong constraints on how we must fill the ``desert''
    of particles below $M_{G}$. }
\item{The magnitude of the singlet $\mu S S$    mass.}
\item{The presence of a nonzero $\nu S$ entry in Eq.~(\ref{generalM})}, proportional to the VEV of the
L-R partner of 
$\chi_{R}$, namely $\chi_{L} \equiv (1,2,1,+1)$.
\end{itemize}
Let us now discuss one by one these three points and show that there
indeed exists a supersymmetric SO(10) model that addresses all these
conditions in a satisfactory way and offers a new way to understand
the smallness of neutrino masses.


First note that there are several mechanisms that could be used to get
rid of the SS-term in Eq.~(\ref{generalM}).  For example we can treat
the SO(10) embedding into $E_{6}$ where the fermionic singlet could be
a member of a 27-dimensional irreducible representation with the
familiar $SO(10)\otimes U(1)_X$ decomposition
\begin{equation}
27_{F}=1_{F}^{4}\oplus 16_{F}^{1}\oplus 10_{F}^{-2}
\end{equation}
If at the $E_{6}$-scale there is no $351'$ Higgs representation the
$U(1)$-charge of the $1_{F}1_{F}$ matter bilinear is so large that it
is very hard to saturate it. Thus, as long as the corresponding $U(1)$
is unbroken we have $\mu=0$. Even if we break the $U(1)$ symmetry
at some lower scale it could be rather complicated to generate an
effective $SS$-entry, which brings further suppression, even at
the level of effective operators. From now on we will neglect
$\mu$.

\begin{figure}[htbp]
  \centering
\includegraphics[scale=.2]{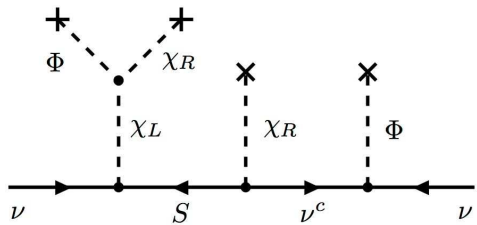} 
  \caption{The proposed supersymmetric \10 seesaw mechanism
    comes from this graph, up to transposition. The neutrino mass is
    suppressed by the unification scale, not by the B-L breaking
    scale, which can be low.}
\label{fig:newseesaw}
  \end{figure}

Now consider the $\nu S$-term.  A typical SO(10) superpotential
contains the following terms,
$$
W\ni M_{16} 16_{H} \overline{16}_{H}+\rho
16_{H}16_{H}10_{H}
+ \rm{H.~c}.
$$
The fact that in 'standard' supersymmetric SO(10) models there is a
small induced vacumm expectation value (VEV) generated for the neutral
component of $\chi_{L}=(1,2,+1)_{SM}\in (1,2,1,+1)_{LR}$ of
$\overline{16}_{H}$ once the B-L symmetry is broken can be seen from
the structure of the F-terms. For example $F^{\dagger}_{(1,2,1,\pm
  1)}$ is proportional to
\begin{eqnarray}
& & M_{16}(1,2,1,\mp 1)_{16}+\rho(1,1,2,\mp
1)_{16}(1,2,2,0)_{10}+\ldots{\nonumber} 
\end{eqnarray} 
After giving a nonzero VEV to the $(1,1,2,\mp 1)$ field (that
subsequently breaks B-L) and the traditional doublet pair in $10_{H}$
(to break the SM) the requirements to be in a supersymmetric vaccum
lead to
\begin{equation}
\label{vL}
\vev{\chi_{L}} \equiv v_{L}\simeq \vev{(1,2,1,\mp 1)_{16}}
\simeq \rho \frac{v_{R} v}{M_{16}}
\end{equation}
Therefore, there is a new contribution to (\ref{generalM}) coming from
a term of the type $ F^{ikl}
{{\nu_{L}}^i}^{T}C^{-1}{S_{L}^{k}}\chi_{L}^{l}+h.c.  $ so that the
neutrino mass matrix reads
\begin{equation}
\label{generalM2}
M_{\nu} =
\left(\begin{array}{ccc}
0 & Y v&  F v_{L} \\
Y^{T}v & 0 & \tilde F v_{R}\\
 F^{T}v_{L} & \tilde F^{T} v_{R}& 0
\end{array}\right) 
\end{equation}
instead of Eq.~(\ref{generalM}). Here $\tilde F$ is an independent
combination of the VEVs of the $\chi_{L}'s$, namely, ${\tilde F}^{ij}
v_L= \sum_k F^{ijk} \vev{\chi_L}^k$, while $F^{ij} v_R= \sum_k F^{ijk}
\vev{\chi_R}^k$.  By inserting $v_{L}$ from (\ref{vL}) into
Eq.~(\ref{generalM2}) one sees that the $v_{R}$ scale drops out
completely from the previous formula, leading to
\begin{eqnarray}
M_{\nu} &\simeq& 
\frac{v^2}{M_{G}}
\rho \left[Y( F \tilde F^{-1})^{T}+( F
  \tilde F^{-1}) Y^{T}\right] 
\end{eqnarray} 
so that the neutrino mass is suppressed by $M_{G}$
\emph{irrespective of how low is the B-L breaking scale}.  This is a
key feature of our mechanism, illustrated in Fig.~\ref{fig:newseesaw}.
In contrast to both the canonical seesaw in Eq.~(\ref{eq:nss}) and the
inverse seesaw, Eq.  (\ref{eq:iss}), this new seesaw is linear in the
Dirac Yukawa couplings $Y$.
Note also that, for given $M_{G}$ and $Y$, the scale of neutrino
masses can be adjusted by choosing appropriately the value of the
cubic scalar sector coupling constant $\rho$, as well as the $ F
\tilde F^{-1}$ (the latter tends to 1 if there is just one copy of
$16_H \oplus \overline{16}_H$). Note also that the current mechanism,
apart from being unified, is also quite distinct from the left-right
symmetric attempts in Refs.\cite{Wyler:1983dd,Akhmedov:1995vm}.
  From this argument it follows that the $B-L$ breaking scale can in
  principle be as low as few TeV.

  
  Concerning the stability of the texture zeros at the $\nu^c\nu^c$
  and $SS$ entries in formula (\ref{generalM2}) it can be protected as
  long as the $U(1)_{R}$ and the $U(1)_{X}$ (of $E_{6}\supset
  SO(10)\otimes U(1)_{X}$ in $E_{6}$ inspired setups) are exact.
  Indeed, $U(1)_{X}$ must be broken at the $v_{R}$ scale
  ($16_{H}\oplus \overline{16}_{H}$ always has a $U(1)_{X}$ charge).
  However, the charge of$\chi_{R}$'s is such that the relevant
  operators arise only at higher orders and may be neglected.



Now we turn to gauge coupling unification.
As was shown by Deshpande et al \cite{Deshpande:1992eu} the scale at
which the $SU(2)_{R}\otimes U(1)_{B-L}$ symmetry is broken to
$U(1)_{Y}$ can be arbitrarily low if we populate properly the
``desert'' from $M_{Z}$ to $M_{G}$. In their case this is achieved by
putting {\it three} copies of $(1,1,2,+1)\oplus (1,1,2,-1)$ coming
from $16_{H}\oplus \overline{16}_{H}$ right at the $v_{R}$ scale.
Then the (one-loop) MSSM running of the $\alpha^{-1}_{Y}$ can be
``effectively'' extended above the $v_{R}$ scale
($\alpha^{-1}_{Y}=\frac{3}{5}\alpha^{-1}_{R}+\frac{2}{5}\alpha^{-1}_{B-L}
$) by a conspiracy between the runing of $\alpha^{-1}_{B-L}$ and
$\alpha^{-1}_{R}$. However, such a scheme is rather {\sl ad hoc} as we
need to push three identical copies of a Higgs multiplet to a very low
scale, at odds with the ``minimal fine-tunning".

Here we present a more compelling scheme in which the $SU(2)_R$
breaking scale $V_R$ is separated from the low $U(1)_{B-L}\otimes
U(1)_R$ breaking scale $v_R$, $V_R \gg v_R$ in the chain $
SU(2)_{R}\otimes U(1)_{B-L}\to U(1)_{R}\otimes U(1)_{B-L}\to
U(1)_{Y}$. At each step we assume just those multiplets needed to
break the relevant symmetry.  The first step is achieved by a light
admixture of the $(1,1,3,0)$ multiplets living in $45$ and $210$ while
the second stage is driven by the light component of the 
$(1,1,+\frac{1}{2},-1)\oplus (1,1,-\frac{1}{2},+1)$ scalars 
(in $SU(3)_{c}\otimes SU(2)_{L}\otimes U(1)_{R}\otimes U(1)_{B-L}$ notation) of
$16_H\oplus\overline{16}_H$.
 Note that to allow for such a L-R asymmetric setup the D-parity of SO(10) must be broken.

 
 Let us further specify the ingredients of our supersymmetric SO(10)
 model needed to implement the mechanism described above.  As usual we
 use three copies of $16_{F}^{i}$ to accomodate the Standard Model
 (SM) fermions and for each of them we add a singlet fermion
 $1_{F}^{i}$ to play the role of $S_{L}$.  A realistic fermionic
 spectrum requires more than one copy of $10_{H}$ Higgs multiplet.
 Moreover we assume one (or more) copy of $16_{H} \oplus
 \overline{16}_{H}$ to implement our new supersymmetric seesaw
 mechanism.  To prevent fast proton decay via dimension 4 operators,
 we assign the matter fermions in $16_{F}$ and $1_{F}$ with a discrete
 matter parity that forbids the mixing of $16_{F}$ and $16_{H}$.
Finally, we add a $45_{H}$ and $210_{H}$ to trigger the proper
symmetry breaking pattern with no D-parity below the GUT
scale~\cite{He:1990jw,He:1989rb,Chang:1985zq}.  The SO(10) invariant
Yukawa superpotential then reads
\begin{equation}
W_{Y}=Y_{aij} 16_{F}^{i}16_{F}^{j}10_{H}^{a}+F_{ijk}16_{F}^{i}1_{F}^{j}\overline{16}_{H}^{k}
\end{equation}
We do not impose other discrete symmetries to reduce the number of
parameters that might however be welcome in connection with the
doublet-triplet splitting problem in a more detailed analysis.  The
Higgs superpotential is
\begin{eqnarray}
W_{H} & = & M_{16}^{kl}{16}_{H}^{k}\overline{16}_{H}^{l}+ M_{10}^{ab}{10}_{H}^{a}{10}_{H}^{b}+{\nonumber}\\
&+& M_{45}{45}_{H}45_{H}+ M_{210}{210}_{H}{210}_{H}+{\nonumber}\\
&+ &\rho_{klm} 16_{H}^{k}16_{H}^{l}10_{H}^{m}
+\overline{\rho}_{klm}\overline{16}_{H}^{k}\overline{16}_{H}^{l}10_{H}^{m}+ \\
&+ &\sigma_{kl} 16_{H}^{k}\overline{16}_{H}^{l}45_{H}+ \omega_{kl}
16_{H}^{k}\overline{16}_{H}^{l}210_{H}+{\nonumber}\\
& + & \lambda 45_{H}^{3}+ \kappa 45_{H}^{2}210_{H}+\xi 45_{H}210_{H}^{2}+ \zeta 210_{H}^{3}{\nonumber}
\end{eqnarray}


The components of $210_{H}$ and $45_{H}$ that receive GUT-scale VEVs
and trigger the breaking of $SO(10)$ to $SU(3)_{c}\otimes
SU(2)_{L}\otimes SU(2)_{R}\otimes U(1)_{B-L}$ are (in Pati-Salam
language) $ 210_{H}\ni(15,1,1)\oplus(1,1,1)$ and $ 45_{H}\ni(15,1,1)$
As shown in \cite{Chang:1985zq,He:1989rb,He:1990jw} this pattern can
accommodate the desired D-parity breaking allowing for an intermediate
L-R symmetric group with an asymmetric particle content, leading to
distinct $g_{L}$ and $g_{R}$ below $M_{G}$. The subsequent
$SU(2)_{R}\to U(1)_{R}$ breaking at $V_R$ is induced by the VEV of a
light superposition of $(1,1,3,0)_{210}$ and $(1,1,3,0)_{45}$ that can
mix below the GUT scale. Next, the $U(1)_{R}\otimes U(1)_{B-L}$ is
broken down at $v_R$ by the VEVs of the light component of type
$(1,1,+\frac{1}{2},-1)\oplus (1,1,-\frac{1}{2},+1)$ coming from
$(1,1,2,-1) \oplus(1,1,2,1)$ of $16_{H}\oplus \overline{16}_{H}$.  The
final SM breaking step is as usual provided by the VEVs of the
$(1,2,2,0)$ bidoublet components.  Note that unlike the example given
in \cite{Deshpande:1992eu} there is no artificial redundancy in the
number of light states living at intermediate scales.


Let us finally inspect the one-loop gauge coupling unification.
Using the normalization convention
$
2\pi t(\mu)=\ln (\mu/M_{Z})
$ we have (for $M_A < M_B$)
$$
\alpha_{i}^{-1}(M_{A})=\alpha_{i}^{-1}(M_{B})+b_{i}(t_{B}-t_{A})
$$
in the ranges $[M_{Z},M_{S}]$, $[M_{S}, v_{R}]$ and $[V_{R},
M_{G}]$, $M_{S}$ is the SUSY breaking scale taken to be at the $\sim$
1 TeV range.  Between $v_{R}$ and $V_{R}$ the two $U(1)$ factors mix
and the running of $\alpha^{-1}_{R}$ and $\alpha^{-1}_{B-L}$ requires
separate treatment.  The Cartans obey the traditional formula (with
"physically" normalized $B-L$ and $Y_{W}$) $ {Y_{W}}=2
T_{3}^{R}+({B-L}) $.  Note that the SO(10) normalization of $b_{B-L}$
is $ b'_{B-L}=\frac{3}{8} b_{B-L} $.

Once the $D$-parity is broken below $M_{G}$ we have $g_{L}\neq g_{R}$.
The Higgs sector in the stage down to $V_{R}$ is as follows: $1\times
(1,1,3,0)$, $1\times (1,1,2,+1)\oplus (1,1,2,-1)$ and $1\times
(1,2,2,0)$. This gives rise to the $b$-coefficients $b_{3}=-3$,
$b_{L}=1$, $b_{R}=4$ and $b_{B-L}=20$.

At the subsequent stage from $V_{R}$ to $v_{R}$ we keep only the weak
scale bidoublet $(1,2,2,0)$ (that below $V_{R}$ splits into a pair of
L-doublets with the quantum numbers $(1,2,+\frac{1}{2},0)\oplus
(1,2,-\frac{1}{2},0)$ under the 
$SU(3)_{c}$ $\otimes$ $SU(2)_{L}$ $\otimes$ $U(1)_{R}$ $\otimes$
$U(1)_{B-L}$ group) and a part of $(1,1,2,+1)\oplus (1,1,2,-1)$ that
is needed to break $U(1)_{R}\otimes U(1)_{B-L}$ to $U(1)_{Y}$ - a pair
of the $\chi_{R}$ fields $(1,1,+\frac{1}{2},-1)\oplus
(1,1,-\frac{1}{2},+1)$.  Since these fields are neutral with respect
to all SM charges the position of the $v_{R}$ scale does not affect
the running of the ``effective $\alpha^{-1}_{1}$" (given by the
appropriate matching condition) and the only effects arise from the
absence of the righthanded $W$-bosons at this stage.  Using the
$SU(2)_{R}$ normalization of the $U(1)_{R}$ charge the matching
condition at $V_{R}$ is trivial.  The relevant b-coefficients of
$SU(3)_{c}\otimes SU(2)_{L}$ and the matrix of anomalous dimensions of
the mixed $U(1)_{R}\otimes U(1)_{B-L}$ couplings are $b_{3}=-3$,
$b_{L}=1$ and
\begin{equation}
\left(\begin{array}{cccc}
\gamma_{11} & \gamma_{12}\\
\gamma_{21} &\gamma_{22}
\end{array}\right)
=
\left(\begin{array}{cccc}
\frac{15}{2} & -1 \\
-1 & 18
\end{array}\right).
\end{equation}
Below the $v_{R}$ scale the model is the ordinary MSSM with the
b-coefficients $b_{3}=-3$, $b_{L}=1$ and $b_{Y}=33/5$ and finally, the
b-coefficients for the SM stage below the $M_{S}$ scale are
$b_{3}=-7$, $b_{L}=-3$ and $b_{Y}=21/5$.  The $v_{R}$-scale matching
condition reads $ \alpha^{-1}_{Y}(v_{R})=\frac{3}{5}
\alpha^{-1}_{R}(v_{R})+\frac{2}{5} \alpha^{-1}_{(B-L)'}(v_{R}) $.
Recalling that $\alpha_1^{-1}(M_Z) =\frac{3}{5}(1-\sin^2 \!
\theta_W)\alpha ^{-1}(M_Z)$ and $\alpha_2^{-1}(M_Z) =\sin^2 \!
\theta_W \alpha ^{-1}(M_Z)$ the initial condition (for central values
of the input parameters) is $\alpha^{-1}_{1}(M_{Z})\doteq 59.38$,
$\alpha^{-1}_{2}(M_{Z})\doteq 29.93$ and $\alpha^{-1}_{3}(M_{Z})\doteq
8.47$~\cite{Eidelman:2004wy}.

\begin{figure}[htbp]
  \centering
\includegraphics[height=4.4cm,width=.9\linewidth]{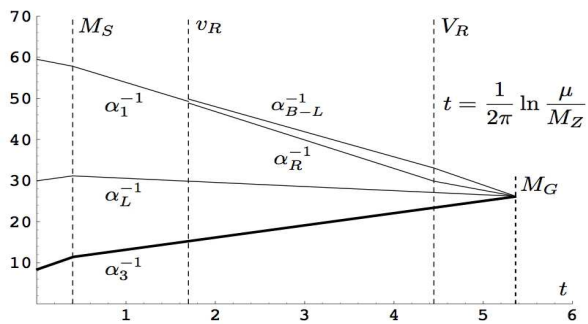}
\caption{\label{running1}
  The one-loop gauge coupling unification in the model described in
  the text.  The D-parity is broken at $M_G$ and the intermediate
  scales $V_R$ and $v_R$ correspond to the $SU(2)_R\to U(1)_R$ and
  $U(1)_R\otimes U(1)_{B-L}\to U(1)_Y$ breaking respectively.}
   \end{figure}

\begin{figure}[htbp]
  \centering
 \includegraphics[height=4.4cm,width=.9\linewidth]{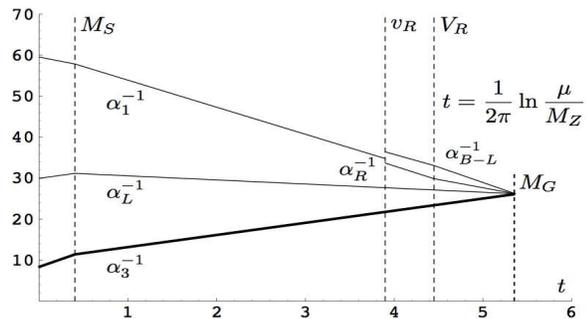}
\caption{\label{running2}
  Same as in Fig.~\ref{running1} for the case of higher $v_R$.  As
  expected, the prediction for $\alpha_1^{-1}(M_Z)$ does not depend on
  the position of the $v_{R}$ scale.}
  \end{figure}

Inspecting the results of the numerical analysis (Figs.\ref{running1}
and \ref{running2}) one shows that, indeed, the $v_{R}$ scale does not
affect the predicted value of $\alpha_1^{-1}(M_Z)$. Its value remains
essentially free at one-loop level. Thus, the unification pattern is
fixed entirely by the interplay of $M_{S}$ and $V_{R}$. The lower
bound $V_{R}\gtrsim 10^{14}$ GeV is consistent with the ``standard"
minimally fine-tuned SUSY SO(10) behaviour, see for instance
\cite{Bajc:2004xe}.
 
  In summary we have proposed a variant supersymmetric SO(10) seesaw
  mechanism which involves a dynamical scale $v_{R}$ that can be
  rather low, as it is essentially unrestricted both by gauge coupling
  unification and neutrino masses.  The smallness of neutrino masses
  coexists with a light $B-L$ gauge boson, possibly at the TeV scale,
  that can be produced at the Large Hadron Collider, by the Drell-Yan
  process.
  Moreover, the "heavy" neutrinos involved in the seesaw mechanism,
  see Eq.~(\ref{generalM2}), get masses at $v_{R}$ and can therefore
  be sufficiently light as to bring a rich set of phenomenological
  implications.  For example their exchange can induce flavour
  violating processes, like $\mu\to e\gamma$ with potentially very
  large rates, similar to the inverse seesaw model of
  Eq.~(\ref{generalM}))~\cite{bernabeu:1987gr,Ilakovac:1994kj}.  We
  conclude that, within the unified SO(10) seesaw picture, the
  dynamics underlying neutrino masses may have observable effects at
  accelerators and in the flavor sector.

  This work was supported by the Spanish grant BFM2002-00345 and by
  the EC Human Potential Programme RTN network MRTN-CT-2004-503369.
  We thank Rabi Mohapatra and Goran Senjanovi\'c for a useful discussion.
  
  \textsl{Note added}: after completing this work we came across
  related papers by Steve Barr and collaborators, and by Fukuyama et
  al~\cite{Barr:2003nn} who revive the type-III seesaw suggested
  in~\cite{Akhmedov:1995vm} and give it a different theoretical
  context. All of these have indeed elements in common, as well as
  with the early work in Ref.~\cite{mohapatra:1986bd,bernabeu:1987gr}.
  However, the mechanism we now propose differs crucially from all of
  the previous in that our B-L scale can be very low, in contrast to
  that of the other models.  We show how our new and key feature not
  only accounts for the oberved neutrino mass scale, but also fits
  with the gauge unification condition in SO(10).


\end{document}